\documentclass[showpacs,preprint,amsmath,amssymb]{revtex4}
\usepackage{graphicx}
\begin{document}
\title{A simple method to obtain the equilibrium solution of
Wigner-Boltzmann Equation with all higher order quantum
corrections}
\author{Anirban Bose$^1$ and M.S. Janaki$^2$}
\affiliation{ $^1$ Serampore College, Serampore, Hooghly.\\
$^2$Saha Institute of Nuclear Physics, I/AF Bidhannagar,
Calcutta 700 064, India}

\begin{abstract}
A simple method has been introduced to furnish the equilibrium solution of the Wigner equation for all order of the quantum correction. This process builds up a recursion relation involving the coefficients of the different power of the velocity. The technique greatly relies upon the proper guess work of the trial solution and is different from the Wigner's original work. The solution is in a compact exponential form with a polynomial of velocity in the argument and returns the Wigner's form when expansion of the exponential factor is carried out. The study keeps its importance in studying various close as well as open quantum mechanical system. In addition, this solution may be employed to obtain the nonequilibrium one particle wigner distribution in the  relaxation-time approximation and under near-equilibrium conditions.
\end{abstract}
 \pacs{05.30.-d, 05.20.Dd}
\maketitle

\section{Introduction}
In quantum mechanics the direct approach of obtaining the partition function is concerned with the determination of the energy levels by solving the schrodinger equation with proper symmetry condition imposed. Von Neumann \cite{kn:neumann} put forward the method that eventually converted the summation into a phase space integral analogous to Gibbs integral. This process was utilized by Bloch \cite{kn:bloch} and Wigner \cite{kn:wigner}. Wigner, in the semiclassical limit, expanded the expression in the power of Planck constant and got back the Gibbs distribution in the classical case. The main advantage of the phase space probability function is seeded in the ability to produce the quantum mechanical averages which is very much similar to that of
the classical mechanics. Kirkwood \cite{kn:kirkwood} took on the same expansion scheme but applied it on the Bloch equation to generate a recursion relation for the higher order calculation. In addition to that the requirement of symmetry property of the wave function was taken care of. Correction of thermodynamic quantities was performed on the basis of the work of Kirkwood of developing the Slater sum. Uhlenbeck  and Beth \cite{kn:uhlenbeck} worked on the quantum modification of the virial coefficients of gas. Subsequently, de Boer \cite{kn:deboer} calculated other thermodynamic quantities.

On the other hand, the attempt of Green \cite{kn:green} was to obtain the direct solution of
the fundamental Wigner equation. This solution is a power series in $\hbar^{2}$ and it matches well with the results manifested by other techniques. Its applicability in the condensed system and gases reflects the utility of the method.

This remarkable Wigner equation is concerned with the close quantum system.
In the next stage, investigation was carried out in
the direction of probing the open system. In order to do that the Fokker-Planck equation was extended in the quantum regime. Recently \cite{kn:coffey}, the classical Maxwell-Boltzman distribution has been modified to include the quantum corrections to the dissipative non-equilibrium dynamics governing the quantum Brownian motion in the presence of external potential. The result, in the weak coupling and high temperature limit, gives way to a semiclassical master equation for the reduced Wigner function in quasi-phase space.

The approach\cite{kn:wigner2} of Wigner is basically the reformulation of Schr$\ddot{o}$dinger wave mechanics where the state of the system is described by the functions in the configuration space. In spite of its noteworthy presence in the quantum statistical mechanics, it also finds itself applicable in subjects like quantum chemistry and quantum optics.

However, Kirkwood, in his article, correctly pointed out that a recursion formula from the Bloch equation greatly simplifies the higher order calculations than the method employed by Wigner. Wigner also acknowledged this point in his remarkable article.

The motivation of this article lies in the construction of the solution of the Wigner equation with all higher order quantum corrections through a  different simple approach  and that ultimately minimizes the amount of effort previously involved in the wigner's own technique. This process gives rise to solving of a series of simple first order differential equations and that ultimately builds up a recursion relation involving the coefficients of the powers of the velocity. The dissimilarity with the original approach of Wigner is revealed in the complete solution of the Wigner equation. The structural difference is noticeable and the extraction of the Wigner's solution is possible through the expansion of the newly obtained solution. The relevance of this work is intimately linked up with the development of a complete QHD theory\cite{kn:gardner} to probe the semiconductor physics and more accurately describe the equation of state of electron gas\cite{kn:ancona} and build up a generalized diffusion�drift-current equation. The semiclassical approach to quantum Brownian motion based on the mapping of quantum operators in Hilbert space to a classically meaningful representation space, has been constructed on the equilibrium solution of the Wigner equation with higher order corrections. The result of this article may be utilized for the subsequent development of the master equation in quasi-phase space
valid in the weak coupling and high temperature limit, to study the open system where the Wigner-Moyal
equation is coupled to a collision kernel in the form of a Kramers-Moyal expansion.

\section{Solution of The Complete Wigner Equation}
The wigner equation is
\begin{eqnarray}
&&\frac{\partial f}{\partial t}+\frac{p}{m}
\frac{\partial f}{\partial {x}}-\frac{\partial
\phi}{\partial {x}}\frac{\partial f}{\partial
{p}}+ \sum_{j=1}^{\infty}(-1)^{j+1}C_{j}\hbar^{2j}\frac{\partial^{2j+1}
\phi}{\partial x^{2j+1}}
\frac{\partial^{2j+1} f}{\partial p^{2j+1}}=0 \label{v3}\end{eqnarray}

where, $$C_{j}=1/{(2)}^{2j}(2j+1)!$$

The semiclassical equilibrium solution will be determined in a perturbative way. The starting point will be the first equation, where the first term of the infinite series will be retained and the following solution will be obtained.

\begin{equation}W_{2}=\exp(a_{01}+a_{11} \frac{p^{2}}{2m})\end{equation}

where,
\begin{equation}a_{01}=-\beta (\phi -\Lambda\beta\frac{d\phi}{dx}\frac{d\phi}{dx}+3\Lambda\frac{d^{2}\phi}{dx^{2}})\end{equation}
\begin{equation}a_{11}=-\beta \frac{p^{2}}{2m}(1-2 \Lambda\frac{d^{2}\phi}{dx^{2}})\end{equation}

and $$\Lambda=\beta^{2}\hbar^{2}/24m$$
In the next stage, the first two terms of the series will be taken into account and the the solution will be sought of the following form

$$f=W_{2}W_{4}$$

where,

\begin{equation}W_{4}=\exp{[\Lambda^{2}(a_{02}+a_{12}\frac{p^{2}}{2m}+a_{22}{{(\frac{p^{2}}{2m})}^{2}})]}\end{equation}

Inserting the solution in the equation
\begin{eqnarray}
&&\frac{\partial f_{1}}{\partial t}+\frac{p}{m}
\frac{\partial f_{1}}{\partial {x}}-\frac{1}{m}\frac{\partial
\phi}{\partial {x}}\frac{\partial f_{1}}{\partial
{p}}+ C_{1}\hbar^{2}\frac{\partial^3
\phi}{\partial x^{3}}
\frac{\partial^3 f_1}{\partial p^{3}}-C_{2}\hbar^{4}\frac{\partial^5
\phi}{\partial x^{5}}
\frac{\partial^5 f_1}{\partial p^{5}} =0 \label{v3}\end{eqnarray}

and dropping the higher order terms, the following equation is obtained.

\begin{eqnarray}
&&(\frac{p}{m}
\frac{\partial W_{2}}{\partial {x}}-\frac{\partial
\phi}{\partial {x}}\frac{\partial W_{2}}{\partial
{p}}+ C_{1}\hbar^{2}\frac{\partial^3
\phi}{\partial x^{3}}
\frac{\partial^3 W_2}{\partial p^{3}})W_{4}-C_{2}\hbar^{4}\frac{\partial^5
\phi}{\partial x^{5}}
\frac{\partial^5 W_2}{\partial p^{5}}W_{4}+\\
&&(\frac{p}{m}\frac{\partial W_{4}}{\partial {x}}-\frac{\partial
\phi}{\partial {x}}\frac{\partial W_{4}}{\partial
{v}}+ C_{1}\hbar^{2}\frac{\partial^3
\phi}{\partial x^{3}}
\frac{\partial^3 W_4}{\partial p^{3}})W_{2}-C_{2}\hbar^{4}\frac{\partial^5
\phi}{\partial x^{5}}
\frac{\partial^5 W_4}{\partial p^{5}}W_{2} =0 \end{eqnarray}

As $W_{2}$ satisfies the first equation of the perturbative series, the first three terms
of the equation disappears. The last two terms, on the other hand, being the higher order
contribution, can be safely neglected.
The remaining terms
are simplified to the following forms

\begin{equation}\frac{p}{m}\frac{\partial W_{4}}{\partial x}W_{2}=\Lambda^{2}\frac{p}{m}(\frac{\partial a_{02}}{\partial x}+\frac{\partial a_{12}}{\partial x}\frac{p^{2}}{2m}+\frac{\partial a_{22}}{\partial x}{(\frac{p^{2}}{2m})}^{2})W_{4}W_{2}\end{equation}

\begin{equation}-\frac{\partial
\phi}{\partial {x}}\frac{\partial W_{4}}{\partial
{p}}W_{2}=-\frac{\partial
\phi}{\partial {x}}\Lambda^{2}(a_{12}\frac{p}{m}+a_{22}{\frac{p^{3}}{m^{2}}})W_{4}W_{2}\end{equation}

\begin{equation}-C_{2}\hbar^{4}\frac{\partial^5
\phi}{\partial x^{5}}
\frac{\partial^5 W_2}{\partial p^{5}}W_{4}=-C_{2}\hbar^{4}\frac{\partial^5
\phi}{\partial x^{5}}[2{(\frac{\mu}{m})}^{2}+9{(\frac{\mu}{m})}^{3}p+6{(\frac{\mu}{m})}^{3}p^{2}+3{(\frac{\mu}{m})}^{4}p^{3}+{(\frac{\mu}{m})}^{4}p^{4}]W_{2}W_{4}\end{equation}

Collecting the coefficients of different powers of v and separately equating them to zero, the following equations emerge out

$p^{5}\longrightarrow$
\begin{equation}\frac{\partial a_{22}}{\partial x}\frac{\Lambda^{2}}{4m^{3}}=C_{2}{\hbar}^{4}\frac{\partial^5
\phi}{\partial x^{5}}{(\frac{\mu}{m})}^{5}\end{equation}

$p^{3}\longrightarrow$
\begin{equation}\frac{\partial a_{12}}{\partial x}=6\frac{\partial^5
\phi}{\partial x^{5}}+2\frac{\partial
\phi}{\partial {x}}a_{22}\end{equation}

$p\longrightarrow$
\begin{equation}\frac{\partial a_{02}}{\partial x}=\frac{\partial
\phi}{\partial x}a_{12}-\frac{9}{2\beta}\frac{\partial^{5}
\phi^{}}{\partial {x}^{5}}\end{equation}

From  (12)
\begin{equation}a_{22}=-\frac{6}{5}\beta\frac{\partial^4
\phi}{\partial x^{4}}\end{equation}

\begin{equation}a_{12}=6\frac{\partial^4
\phi}{\partial x^{4}}-\frac{12\beta}{5}(\frac{\partial
\phi}{\partial x}\frac{\partial^3
\phi}{\partial x^{3}}-\frac{1}{2}\frac{\partial^2
\phi}{\partial x^{2}}\frac{\partial^2
\phi}{\partial x^{2}})\end{equation}

\begin{equation}a_{02}=-\frac{9}{2\beta}\frac{\partial^4
\phi}{\partial x^{4}}+6(\frac{\partial
\phi}{\partial x}\frac{\partial^3
\phi}{\partial x^{3}}-\frac{1}{2}\frac{\partial^2
\phi}{\partial x^{2}}\frac{\partial^2
\phi}{\partial x^{2}})-\frac{12\beta}{5}\int (\frac{\partial
\phi}{\partial x}\frac{\partial
\phi}{\partial x}\frac{\partial^3
\phi}{\partial x^{3}}-\frac{1}{2}\frac{\partial
\phi}{\partial x}\frac{\partial^2
\phi}{\partial x^{2}}\frac{\partial^2
\phi}{\partial x^{2}})\end{equation}
%
%
%
This process can be continued for the remaining terms of the infinite series and the final solution will be obtained as

$$f=\prod_{j=1}^{\infty}W_{2j}$$

For j=1, we have already obtained $W_{2}$

For $j>1$,
$$W_{2j}=\exp (U_{2j})$$
where
$$U_{2j}=\sum_{i=0}^{j}\hbar^{2j} a_{ij}({\frac{p}{\sqrt{2m}}})^{2i}$$
and the following set of equations stand out

\begin{equation}\frac{1}{m}\frac{\Lambda^{j}}{(2m)^{i}}\frac{\partial a_{ij}}{\partial x}=2(i+1)a_{(i+1)j}\frac{\Lambda^{j}}{(2m)^{i+1}}\frac{\partial \phi}{\partial x}+b_{ij}\end{equation}

where,

\begin{equation}b_{ij}=-(\hbar)^{2j}\frac{\partial^{2j+1} \phi}{\partial x^{2j+1}}(\frac{2m}{\beta})^{-(2j+1)}D_{i}C_{j}\end{equation}
where, $D_{i}=$ coefficient of $p^{2i+1}$ of the Hermite polynomial $H_{2j+1}(\sqrt{\frac{\beta}{2m}}p)$

This equations are true for i=0 to i=j-1

For i=j,

\begin{equation}\frac{1}{m}(\frac{\Lambda}{2m})^{j}\frac{\partial a_{ij}}{\partial x}=-\hbar^{2j}\frac{\partial^{2j+1} \phi}{\partial x^{2j+1}}(\frac{2m}{\beta})^{-(2j+1)}C_{j}D_{j}\end{equation}
where

This is a first order equation and can be easily solved to obtain the value of $a_{jj}$. Using this result, all the equations of this group can be successively solved to find out the remaining coefficients.
\section{Conclusion}

In this article the solution has been obtained
as a exponential function whose argument is a
polynomial of even power of momentum and their
coefficients are the functions of the potential.
This particular form automatically preserves
the positive nature of the distribution function.
If the terms involving the Planck constant are
switched off, the classical maxwell-boltzmann
distribution function is retained. The study of
the exponential function leads to some specific
conclusion regarding the pattern of the distribution
function. First of all, it is observed that as the
order of the correction increases, the highest power
of the momentum goes up by two orders of magnitude from
its previous order solution.
The coefficients of the highest power of the momentum, for
a particular order of correction, contain the highest
order derivatives of the potential function.
This exponential
solution when expanded in the power of Planck constant returns
back the form obtained by Wigner in his famous paper. From
the set of the general equations it is evident that the
determination of the coefficient of the highest power
of momentum is trivial. In the next stage, the equation
involves this coefficient of the highest power of momentum
and the determination of the other coefficient of the series
amounts to solving few first order differential equations
where the right hand side are the function of the
potential. The process has been demonstrated to find
out the second order correction to the distribution
function. It seems to be simple and
transparent and determining the coefficients becomes
straight forward. If the equation of a particular
order is looked into, a regular pattern floats up. In this connection, the
role of the product form is  of immense importance. The
solution of a particular order is the product of the
solution of the previous order and a new correction
term. The differential operator works on
the solution and the product rule of differentiation
applies.

\begin{equation}\frac{\partial^{m}(f_{2n-2}Y)}{\partial x^{m}}=\frac{\partial^{m}f_{2n-2}}{\partial x^{m}}Y+A_{1}\frac{\partial^{m-1}f_{2n-2}}{\partial x^{m-1}}\frac{\partial Y}{\partial x}+...+\frac{\partial^{m}Y}{\partial x^{m}}f_{2n-2}\end{equation}
where

\begin{equation}f_{2n-2}=\prod_{i=1}^{n-1}W_{2i}\end{equation}

and $Y=W_{2n}$ is the correction factor for a particular order.
When differentiated, the $\hbar^{2n}$ in the  exponential structure of Y brings more $\hbar$ and that ultimately leads to  higher order terms. Consequently, all the members except the first one may be safely dropped off. A $\hbar$ free term in the exponential structure of W is responsible to keep the first term of the series alive. Few of them combine to produce the previous order equation and leads to the vanishing of those part. Finally, the equation shapes up in the following form.
\begin{eqnarray}
&&(\frac{ p}{m}
\frac{\partial W_{2n}}{\partial {x}}-\frac{\partial
\phi}{\partial {x}}\frac{\partial W_{2n}}{\partial
{p}})W_{2n-2}+(-1)^{n+1}C_{n}\hbar^{2n}\frac{\partial^{2n+1}
\phi}{\partial x^{2n+1}}
\frac{\partial^{2n+1} W_{2n-2}}{\partial p^{2n+1}}W_{2n}=0\end{eqnarray}

This recursion relation links a particular order solution
to that of the previous order and has been employed to derive the set
of the equations involving the coefficients of the
momentum function. Kirkwood pointed out that a recursion relation in the Bloch formalism simplifies the derivation of higher order correction and makes it more effective than the Wigner's. Now this recursion relation obtained in this article does the same for the Wigner formalism and contribute significantly to the effectiveness of the process.

Finally, in this article a simple method to obtain the
equilibrium solution of the complete Wigner equation
for all order of the quantum correction has been
discussed. This process is based on the correct guess
work of the trial solution and that effectively reduces
the task to solving few first order simple differential
equations. This solution will be useful to study a variety
of closed quantum systems. Moreover, in the weak coupling
limit of the particle heat bath interaction, this technique can be employed for the implementation of
the  Fokker-Planck equation in the quantum regime and
ultimately facilitates the probing of the open quantum
systems. This solution can
be useful for the derivation of the equation of state
of electron gas in the semiconductor device when the
quantum correction is retained to all orders. In addition, in
the near equilibrium state the non-equilibrium wigner
distribution function is obtained from the equilibrium
solution and that proves to be helpful in computing the
current density of the system. In semiconductor physics
the dependence of gradient of
density was included in the equation of state for
the electron gas and that ultimately led to a generalized
diffusion �drift-current equation. The underlying
microscopic nature was investigated through the
semiclassical wigner equation to the lowest order of quantum
correction \cite{kn:ancona}. The result obtained in
this article could be useful for the subsequent
development of a more complete transport equation when the physical condition does not allow the higher order terms to be turned off.


The usefulness of this new distribution function with all higher order
correction is also reflected for the construction of a more complete QHD empowered with the investigation capability
of the low temperature regime where the higher order terms
is of significant importance. As the temperature
of the system is lowered, the contributions from
higher correction terms gradually
creep up. Consequently, the investigation of such
low temperature system with the incomplete
solution of the wigner equation will be ended
up with the imperfect information about the
thermodynamic as well as time evolution property of those
systems. Therefore, the result can serve as a new inexpensive approach to meet the necessity of a complete solution with all higher
order correction and that will be
the true building block for the probing of
the low temperature system where the
semiclassical approach is still meaningful
but the higher order terms can not be put off.

\end{document}